\documentclass[prl,twocolumn,showpacs,amsfonts,amsmath,floatfix]{revtex4}
\usepackage{graphicx}
\newcommand{\Journal}[4]{#1 {\bf #2}, #3 (#4)}
\newcommand{\PRL}{Phys. Rev. Lett.}

\newcommand{\Science}{Science}

\begin{document}
\title{Tonks-Girardeau and Super Tonks-Girardeau States of a Trapped 1D Spinor Bose Gas}
\author{M. D. Girardeau}
\email{girardeau@optics.arizona.edu}
\affiliation{College of Optical Sciences, University of Arizona, Tucson, AZ 85721, USA}
\date{\today}
\begin{abstract}
A harmonically trapped ultracold 1D spin-1 Bose gas with strongly repulsive or attractive 1D even-wave interactions induced by a 3D Feshbach resonance is studied. The exact ground state, a hybrid of Tonks-Girardeau (TG) and ideal Fermi gases, 
is constructed in the TG limit of infinite even-wave repulsion by a spinor Fermi-Bose mapping to a spinless ideal Fermi gas. 
It is then shown that in the limit of infinite even-wave attraction this same state remains an exact many-body eigenstate, 
now highly excited relative to the collapsed generalized McGuire cluster ground state, showing that the hybrid TG state is completely
stable against collapse to this cluster ground state under a sudden switch from infinite repulsion to infinite attraction. 
It is shown to be the TG limit of a hybrid super Tonks-Girardeau (STG) state which is metastable under a sudden switch from finite but very strong repulsion to finite but very strong attraction. It should be possible to create it experimentally by a sudden switch 
from strongly repulsive to strongly attractive interaction, as in the recent Innsbruck experiment on a spin-polarized
bosonic STG gas. In the case of strong attraction there should also exist another STG state of much lower energy, 
consisting of strongly bound dimers, a bosonic analog of a recently predicted STG state which is an ultracold gas of strongly bound bosonic dimers of fermionic atoms, but it is shown that this STG state cannot be created by such a switch from strong repulsion
to strong attraction.
\end{abstract}
\maketitle
If an ultracold atomic vapor is confined in a de Broglie wave guide with
transverse trapping so tight and temperature so low that the transverse
vibrational excitation quantum is larger than available
longitudinal zero point and thermal energies, the effective dynamics becomes
one-dimensional (1D) \cite{Ols98,PetShlWal00}. 3D Feshbach resonances \cite{Rob01}
allow tuning to the neighborhood of
1D confinement-induced resonances \cite{Ols98,BerMooOls03} where the 1D interaction is very
strong, leading to strong short-range correlations, breakdown of
effective-field theories, and emergence of highly-correlated $N$-body
ground states. In the case of spinless or spin-polarized bosons with zero-range Lieb-Liniger (LL) \cite{LieLin63} delta function 
repulsion $g_B\delta(x_j-x_\ell)$ with coupling constant $g_B\to +\infty$, the
Tonks-Girardeau (TG) gas, the exact $N$-body ground state was determined in 1960 by a
Fermi-Bose (FB) mapping to an ideal Fermi gas \cite{Gir60}, leading to
``fermionization'' of many properties of this Bose system, as recently
confirmed experimentally \cite{Par04,Kin04}. 

It was predicted several years ago \cite{AstBluGioGra04,BatBorGuaOel05,AstBorCasGio05} 
that if $g_B$ is large and negative, there exists a highly excited gaseous 
state known as the super Tonks-Girardeau (STG) gas which is  metastable against collapse to the McGuire cluster ground state 
\cite{McG64}, and this state was recently created by the Innsbruck group \cite{Haletal09}
by suddenly switching the interaction from strongly repulsive to strongly attractive by passing through the 1D confinement-induced
resonance induced by a 3D s-wave Feshbach resonance. This initially surprising metastability was recently shown \cite{GirAst09}
to result from the fact that after a sudden switch from infinite repulsion ($g_B=+\infty$) to infinite attraction ($g_B=-\infty$),
the $g_B=+\infty$ TG ground state remains an exact energy eigenstate, now highly excited relative to the collapsed McGuire
cluster ground state, and is therefore stationary and hence completely stable against collapse. Then by continuity,
the system should be metastable under a sudden switch from strong repulsion to strong attraction as in the Innsbruck experiment
\cite{Haletal09}.

A generalization of the FB mapping 
to a 1D spinor Fermi gas is relevant to recent predictions of super Tonks-Girardeau (STG)
states in such a system \cite{CheGuaYinGuaBat09,Gir10}. The model consists of an ultracold 1D gas of
fermionic atoms with very strong attractive interactions in an effectively 1D trap, in two different hyperfine states. Similar STG states should also occur in a 1D spinor Bose gas consisting of bosonic atoms in three hyperfine states conveniently labeled by spin z-component labels $\sigma=-1,0,1$ of spin-1 atoms, but have not been studied so far,
although the case of strong repulsion (TG regime) has been studied in detail by a spinor generalization of the FB mapping
\cite{Deuetal08}. 
The Hamiltonian used there included the kinetic energy, a spin-independent harmonic trapping potential,
a zero-range LL interaction \cite{LieLin63}, and a spin-spin interaction, but simplifies in the TG regime where the spin-spin interaction
is negligible. Then the Hamiltonian is of the same form as that in 
\cite{CheGuaYinGuaBat09,Gir10}, consisting of the kinetic energy, a spin-independent harmonic trap potential, and a
spin-independent LL interaction:
\begin{equation}\label{H}
\hat{H}_{\text{B}}=\sum_{j=1}^N\left(-\frac{\hbar^2}{2m}\frac{\partial^2}{\partial x_j^2}+\frac{m\omega^2}{2}x_j^2\right)
+g_{\text{B}}\hspace{-0.3cm}\sum_{1\le j<\ell\le N}\delta(x_j-x_\ell)\ .
\end{equation}
The case relevant to
generation of a STG state is that of strong s-wave scattering due to a 3D s-wave Feshbach resonance, which leads in 1D to a 
confinement-induced even-wave resonance \cite{Ols98,BerMooOls03} and a LL delta function interaction \cite{LieLin63}. It acts
between atoms $j$ and $\ell$ only if the relative wave function is symmetric under spatial exchange $(x_j\leftrightarrow x_\ell)$, in which case
bosonic symmetry under combined space-spin exchange $(x_j,\sigma_j)\leftrightarrow(x_\ell,\sigma_\ell)$ demands that 
the relative wave function be symmetric under spin exchange $\sigma_j\leftrightarrow\sigma_\ell$. 
On the other hand, if the relative wave function is antisymmetric under spatial exchange, then the interaction term in 
Eq. (\ref{H}) is cancelled by the node at $x_j=x_\ell$.

If $g_{\text{B}}$ is large and negative, two quite different STG states can occur, as in \cite{Gir10}. 
One is a hybrid between an STG gas with strong attractions between atoms in spin-symmetric pair states and an ideal Fermi gas with no
interactions, the STG component being an exact analog of the STG state
in an ultracold 1D Bose gas predicted in \cite{AstBluGioGra04,BatBorGuaOel05,AstBorCasGio05} and 
recently created by the Innsbruck group \cite{Haletal09}. 
It should be possible to create it experimentally by a sudden switch of the 
interaction from strongly repulsive to strongly attractive, as in \cite{Haletal09}.
The other STG state is a trapped analog of a recently predicted STG state in the spinor Fermi gas \cite{CheGuaYinGuaBat09},
an ultracold gas of strongly bound 
boson dimers which behave as bosons with a strongly attractive boson-boson interaction, leading to STG behavior.

Consider first the case $N=2$. The analysis closely parallels that for the spinless Bose gas \cite{GirAst09} and the spinor Fermi gas 
\cite {Gir10}. The LL interaction is
$g_{B}\delta(x_1-x_2)$, the harmonic trap potential is $m\omega^{2}(x_1^{2}+x_2^{2})/2$, the wave function for the center of mass (c.m.) coordinate $X=(x_1+x_2)/2$ is  $\psi_{\text{c.m.}}=\exp[-X^2/x_{\text{osc}}^2]$ where $x_{\text{osc}}=\sqrt{\hbar/m\omega}$, 
and its energy is $E_{\text{c.m.}}=\hbar\omega/2$ assuming that the c.m. mode is unexcited. Bosonic symmetry requires that the 
relative wave function $\psi_{\text{rel}}(x_1,\sigma_1;x_2,\sigma_2)$ be symmetric under conbined space-spin exchange.
Since the Hamiltonian is spin-independent, $\psi_{\text{rel}}$ factorizes into the product of a spatial wave function 
$\phi(x)$, with $x=x_{1}-x_{2}$, and a spin-pair wave function $\chi(\sigma_1,\sigma_2)$, where either both $\phi$ 
and $\chi$ are symmetric or else both are antisymmetric. If they are antisymmetric then $\phi$ vanishes at contact $x=0$,
killing the LL interaction $g_{B}\delta(x_1-x_2)$, so that $\phi$ reduces to a trapped harmonic oscillator eigenstate odd in $x$, the product of the ground state Gaussian and a Hermite polynomial of odd order. 
In the symmetric case $\phi$ is identical with the solution for spinless bosons
\cite{Fra03,TemSolSch08,GirAst09,Gir10}. These solutions are analytic continuations of the Hermite-Gaussians to 
nonintegral quantum number $\nu$, and are parabolic cylinder function $D_{\nu}(q)$ where $x=qx_{\text{osc}}$ and the allowed values of 
$\nu$ are solutions of a transcendental equation
$\Gamma(\frac{1}{2}-\frac{1}{2}\nu)/\Gamma(-\frac{1}{2}\nu)=-\lambda$
where $\lambda$ is a dimensionless coupling constant $\lambda=g_B/(2^{3/2}\hbar\omega x_{\text{osc}})$ \cite{Fra03}. This gives the solution
for $q\ge 0$, while for $q<0$ it is $D_{\nu}(|q|)$, since $\phi(x)$ is even. For $\lambda_B\to+\infty$ the solution reduces to
the first excited harmonic oscillator state $\phi(x)=|q|e^{-q^2/2}$ \cite{GirWriTri01}, the usual TG gas ground state with a cusp at
$q=0$ due to the point hard core interaction. For $\lambda\to -\infty$ this wave function is still an exact energy eigenstate,
but it is highly excited, the ground state being a collapsed state which is an analog, for the trapped system, of
McGuire's cluster state \cite{McG64}. It is an even solution also expressible
in terms of a $D_\nu$, but one whose energy approaches $-\infty$ as $g_B\to -\infty$ ($a_{\text{1D}}\to 0+$); see Fig. 5 of \cite{Fra03}.
For $a_{\text{1D}}\to 0+$ it is well approximated by
$\psi_{B0}\approx\exp(-|x_1-x_2|/a_{\text{1D}})\exp[-(x_1^2+x_2^2)/2x_{\text{osc}}^2]$ where $a_{\text{1D}}\ge 0$ is the 1D 
scattering length. $a_{\text{1D}}$ vanishes as $|\lambda|\to\infty$ and increases with decreasing $|\lambda|$ 
\cite{Ols98,Fra03,TemSolSch08}, and at the same time the node moves from the origin to a position $x_{\text{node}}$
which is close to $a_{\text{1D}}$ when $|\lambda|\gg 1$ but slightly smaller, the explicit expression being given in \cite{GirAst09}. 

An explicit formula for an exact $N=2$ ground state in the TG limit $\lambda_B\to +\infty$ can be given in terms of a mapping function 
$\alpha(x_1,\sigma_1;x_2,\sigma_2)$ which is space-spin antisymmetric and everywhere $\pm 1$, 
which generates the strongly-interacting ground state $\psi_{B0}$
from the spinless ideal Fermi gas ground state $\psi_{\text{ideal},0}=H_1(q_1-q_2)e^{-(x_1^2+x_2^2)/(2x_{osc}^2})$
where $H_1$ is the Hermite polynomial of order one,
the lowest antisymmetric state. The desired representation is 
$\psi_{B0}=\alpha(x_1,\sigma_1;x_2,\sigma_2)\psi_{\text{ideal},0}(x_1,x_2)$ where
\begin{eqnarray}\label{N=2ground}
\alpha(x_1,\sigma_1&;&x_2,\sigma_2)=\sum_{s=-1}^1\delta_{\sigma_1,s}\delta_{\sigma_2,s}\text{sgn}(x_1-x_2)\nonumber\\
&+&\hspace{-0.7cm}\sum_{-1\le s_1<s_2\le 1}(\delta_{\sigma_1,s_1}\delta_{\sigma_2,s_2}-\delta_{\sigma_1,s_2}\delta_{\sigma_2,s_1})\ .
\end{eqnarray}
The signum function $\text{sgn}(x_1-x_2)$ introduces the required cusps at contact $x_1=x_2$ due to the zero-diameter hard-core
interaction for space-even, spin-even scattering, which are absent in the space-odd, spin-odd part of the wave function; note
that $\text{sgn}(x_1-x_2)\psi_{\text{ideal},0}(x_1,x_2)$ is space-even and $\psi_{\text{ideal},0}(x_1,x_2)$ is space-odd. 
$\psi_{B0}$ is a hybrid TG-ideal Fermi gas state. Since the Hamiltonian is spin-independent, the ground state is $3^2=9$-fold 
degenerate, so this $\psi_{B0}$ is only one member of the degenerate ground manifold. One can obtain others by changing the signs
of various terms in (\ref{N=2ground}). The arbitrary overall sign can be determined by the $s=-1$ term in the first summation,
and changing the sign of either of the two others gives two additional states, for a total of $3$. Then changing the sign of 
any of the three terms in the 
second summation multiplies this by $3$, giving an overall total of $9$. These are linearly independent although not orthonormal, 
and span the $9$-dimensional ground manifold. 
In the absence of spin-spin interaction or an external magnetic field, these states are experimentally indistinguishable; they are chosen here for easy generation by mapping from the spinless
ideal Fermi gas, rather than by requiring them to be simultaneous eigenstates of $\hat{S}^2$ and $\hat{S}_z$. 

Generalization to $N>2$ is straightforward. Denote the $N$-atom wave functions by $\psi_B(x_1,\sigma_1;\cdots;x_N,\sigma_N)$. 
Consider first the case $\lambda\to+\infty$. This can be generated from an $N$-particle 1D spinless ideal Fermi gas state 
$\psi_\text{ideal}$
by a space-spin mapping function $M$ which is the product of two-particle mapping functions (\ref{N=2ground}) over all pairs:
\begin{equation}\label{TG-ideal Fermi}
\psi_B=M(x_1,\sigma_1;\cdots;x_N,\sigma_N)\psi_\text{ideal}
\end{equation}
where the spin-dependent Fermi-Bose mapping function $M$ is
\begin{equation}
M(x_1,\sigma_1;\cdots;x_N,\sigma_N)=\prod_{1\le j<\ell\le N}\alpha(x_j,\sigma_j;x_\ell,\sigma_\ell)
\end{equation}
and $\psi_\text{ideal}$ is an energy eigenstate of the trapped 1D ideal gas of ``spinless" fermions, a Slater
determinant of $N$ different harmonic oscillator orbitals. Starting from this ideal Fermi gas, this mapping generates the required
TG-gas cusps in space-symmetric, spin-symmetric scattering channels of the strongly interacting spinor Bose gas, but no
interaction in space-antisymmetric, spin-antisymmetric channels. $M$ is space-spin antisymmetric \cite{Note0} and everywhere $\pm 1$,
hence self-inverse. 
The Hamiltonian commutes with $M$ when all $|x_{j\ell}|>0$ where $x_{j\ell}=x_j-x_{\ell}$, 
and since the interaction term in (\ref{H}) vanishes there $\psi_B$ 
is an energy eigenstate with the same eigenvalue as $\psi_\text{ideal}$. Here we are particularly interested in the  
ground state $\psi_{B0}$, which is mapped from the ground state $\psi_{\text{ideal}0}$ of the trapped ideal Fermi gas, a 
Slater determinant of the lowest $N$ single-particle eigenfunctions $\phi_n$ of the harmonic oscillator (HO): 
\begin{equation}
\psi_{\text{ideal}0}(x_{1},\cdots,x_{N})=\frac{1}{\sqrt{N!}}
\det_{(n,j)=(0,1)}^{(N-1,N)}\phi_{n}(x_{j})\ .
\end{equation}
The HO orbitals are
\begin{equation}
\phi_{n}(x)= \frac{1}
{\pi^{1/4}x_{osc}^{1/2}\sqrt{2^{n}n!}}e^{-Q^{2}/2}H_{n}(Q)
\end{equation}
with $H_n(Q)$ the Hermite polynomials and $Q=x/x_{osc}$. The ground state is a van der Monde determinant \cite{GirWriTri01}
\begin{eqnarray}
\det_{(n,j)=(0,1)}^{(N-1,N)}H_{n}(x_{j})
& = & 2^{N(N-1)/2}\det_{(n,j)=(0,1)}^{(N-1,N)}(x_{j})^{n} \nonumber\\
& = & 2^{N(N-1)/2}\prod_{1\le j<\ell\le N}(x_\ell-x_j)\ ,
\end{eqnarray}
yielding an exact analytical expression for the $N$-boson ground state:
\begin{eqnarray}\label{N>2_ground}
&&\psi_{B0}(x_{1},\sigma_1;\cdots;x_{N},\sigma_N)
=C_{N}\left[\prod_{i=1}^{N}e^{-Q_{i}^{2}/2}\right]\nonumber\\
&&\times\prod_{1\le j<\ell\le N}\alpha(x_j,\sigma_j;x_\ell,\sigma_\ell)(x_j-x_\ell)
\end{eqnarray}
with normalization constant
\begin{equation}
C_{N}=2^{N(N-1)/4}\left (\frac{1}{x_{osc}} \right )^{N/2}
\left[N!\prod_{n=0}^{N-1}n!\sqrt{\pi}\right]^{-1/2}\ .
\end{equation}
This is an exact ground state in the TG limit $\lambda=+\infty$. It is one member of the $3^N$-fold
degenerate ground manifold, and other members are generated by changing the signs of various terms in $\alpha$, as in the case
$N=2$. If the product defining $M$ were over nonoverlapping pairs $(1,2)$, $(3,4)$, $\cdots(N-1,N)$ then this would give
$(3^2)^{N/2}=3^N$ linearly independent states, covering the ground manifold, but pair overlap in the product is a complication;
however, it seems reasonable to conclude that the ground manifold is covered. 

Suppose that now $\lambda$ is changed instantaneously to $-\infty$. Then the state (\ref{N>2_ground}) is still an exact energy eigenstate, since $\hat{H}$ commutes with $M$ except at like-spin collision points 
$x_j=x_\ell$, where the wave function vanishes with a cusp. In fact, this is an exact energy
eigenstate in the limit $|\lambda|\to\infty$ even in the dissipative case where $\lambda$ is complex, since the wave function vanishes. 
For $\lambda\ll -1$ such a state is 
highly excited, the much lower ground state being a hybrid of a totally collapsed McGuire cluster state \cite{McG64} for like spins and an ideal Fermi gas ground state for unlike spins, as in \cite{Gir10}.

Suppose that $\lambda$ is switched from large positive to large negative values rapidly enough that the sudden approximation is valid. Then the initial wave function will be nearly equal to the state of Eqs. (\ref{TG-ideal Fermi}) and (\ref{N>2_ground}), and the 
wave function after the switch will be a superposition of all eigenstates of the system with the given negative $\lambda$ value,
$\psi_\lambda(t)=\sum_\alpha\langle\psi_{\lambda\alpha}|\psi_{B0}\rangle\psi_{\lambda\alpha}e^{-iE_{\lambda\alpha}t/\hbar}$.
The dominant term in $\psi_\lambda$ will be an STG state which reduces to (\ref{N>2_ground}) as $\lambda\to -\infty$.
The obvious generalization of the N=2 STG wave function to arbitrary $N$ differs
from Eqs. (\ref{TG-ideal Fermi}) and (\ref{N>2_ground}) through replacement of $|x_j-x_\ell|$ by the $N=2$ solution $D_\nu$ for equal spins, while retaining the mapping from the ideal Fermi gas form for unequal spins:
\begin{equation}\label{psiFnu}
\psi_{F\nu}=\left[\prod_{1\le j<\ell\le N}\beta_\nu(x_j,\sigma_j;x_\ell,\sigma_\ell)\right]
\prod_{j=1}^N\exp\left(-\frac{x_j^2}{2x_{\text{osc}}^2}\right)\
\end{equation}
where
\begin{eqnarray}
&&\beta_\nu(x_j,\sigma_j;x_\ell,\sigma_\ell)=
\sum_{s=-1}^1 \delta_{\sigma_j s}\delta_{\sigma_\ell s}D_\nu(|q_{j\ell}|)e^{q_{j\ell}^{2}/4}\nonumber\\
&&\hspace{.1cm}+\hspace{-.6cm}\sum_{-1\le s_1<s_2\le 1}(\delta_{\sigma_j s_1}\delta_{\sigma_\ell s_2}
-\delta_{\sigma_j s_2}\delta_{\sigma_\ell s_1})(x_j-x_\ell)
\end{eqnarray}
and $q_{j\ell}=(x_j-x_\ell)/x_{\text{osc}}$. It satisfies the contact
conditions exactly. One expects the existence of a highly excited gas-like STG state with nodes only at a nearest neighbor separation
$|x_{j\ell}|=x_{\text{node}}$ when the spins are equal, as in the spinless Bose case \cite{GirAst09},
where $x_{\text{node}}$ increases with decreasing
$|\lambda|$, is very close to $a_{1D}$ for $|\lambda|\gg 1$, and goes to zero along with $a_{1D}$ in the TG limit $|\lambda|\to\infty$.
For all $N\ge 2$ the approximate wave functions (\ref{psiFnu}) have exactly these properties, vanishing only at
$|x_{j\ell}|=x_{\text{node}}$ when the spins are equal, and becoming exact both at the collision points $x_{j\ell}=0$ and when all 
$|x_{j\ell}|\to\infty$.
Hence we expect the unknown exact STG excited state for finite negative $\lambda$ to be well approximated by the state of 
Eq. (\ref{psiFnu}). In the exterior region where all $|x_{j\ell}|\ge x_{\text{node}}$, the wave function of this state is identical
with that of a gas of hard spheres of diameter $x_{\text{node}}$ apart from normalization, and the energies of these two states
are exactly equal. The proof is the same as those given previously for the STG states of spinless bosons \cite{GirAst09} and spinor
fermions \cite{Gir10}.

There also exists a quite different STG state when $N_\uparrow=N_\downarrow=N/2$. In this case it was 
shown recently \cite{CheGuaYinGuaBat09}, 
for a system on a ring with no circumferential trapping potential, that for $-\infty<\lambda\ll -1$ there exists a
gas-like energy eigenstate 
in which all the fermions are tightly bound into $(\uparrow\downarrow)$ dimers which behave as
strongly attracting bosons, forming an exact analog of the untrapped solution for spinless bosons \cite{Chenetal}. 
It is a gas-like STG state lying much lower than the fermionic STG state of Eq. (\ref{psiFnu}) due to the negative binding energy of the dimers, but much higher than a McGuire-ideal Fermi hybrid state similar to that of \cite{Gir10}.
One expects that a similar STG state
exists for the present case of harmonic trapping. In fact, for $N=2$ it has already been given as the $N_\uparrow=N_\downarrow=1$
case of the previously-discussed exact $N=2$ solution, and for larger even $N$ there will be a Bose-like STG state 
similar to that of \cite{Chenetal}. However, it could \emph{not} be created by a sudden switch from a strongly repulsive to strongly attractive interaction as in 
\cite{Haletal09}. In fact, the sudden approximation probability of finding the system in such a state after a switch 
$\lambda\gg 1\to -\lambda\ll -1$ is \emph{exactly zero} in the limit $|\lambda|\to\infty$, since the state (\ref{psiFnu})
is an exact energy eigenstate for both $\lambda=+\infty$ and $\lambda=-\infty$. Then by continuity this probability will be $\ll 1$
after a sudden switch $\lambda\gg 1\to -\lambda\ll -1$.
\begin{acknowledgments}
I thank Hans-Cristoph N\"{a}gerl and Vladimir Yurovsky for stimulating conversations, and Gregory Astrakharchik for comments on this paper. 
It was supported by the U.S. Army Research Laboratory and the U.S. Army Research Office under grant number W911NF-09-1-0228. 
\end{acknowledgments}
\end{document}